**Early stages of GaAs nanowires VLS self-catalyzed growth on silica-terminated silicon substrate: A photoemission study**

L. Fouquat, M. Vettori, C. Botella, A. Benamrouche, J. Penuelas, G. Grenet

Institut des Nanotechnologies de Lyon - Université de Lyon, UMR 5270 - CNRS, Ecole Centrale de Lyon, 36 avenue Guy de Collongue, F-69134 Ecully cedex, France

* To whom correspondence should be addressed. E-mail: jose.penuelas@ec-lyon.fr

**Abstract:**

In this paper the early stages of the self-catalyzed Vapor-Liquid-Solid (VLS) growth of GaAs nanowires on epi-ready Si substrates by Molecular Beam Epitaxy (MBE) are studied. The interaction of Ga nano-droplets (NDs) with the silica overlayer is investigated by X-ray Photoemission Spectroscopy (XPS) and Atomic Force Microscopy (AFM). We show how Ga NDs drill the silica overlayer and make contact with bulk Si to allow GaAs nanowires (NWs) epitaxial growth by studying each of the three steps of the NW growth process sequentially: Ga NDs pre-deposition, post-deposition annealing to reach the NW growth temperature, and finally NW growth itself. The pre-deposition temperature allows to control the density and morphology of the NDs. A high enough annealing temperature enhances the reduction of $SiO_x$ by Ga oxidation and leads to the formation of holes which is seen by a new component in XPS spectra, assumed to be the consequence of the Ga-Si interaction. Finally, these nano-holes act as nucleation sites for the epitaxial GaAs VLS growth. During the GaAs growth, three different chemical environments of Ga are identified in Ga2p core level: metallic Ga due to the droplets at the top of the NWs, Ga oxide in contact with the $SiO_x$ overlayer and Ga arsenide from NWs.

1. **Introduction:**

III-V semiconductor nanowires (NWs) are presently intensively studied because of their unique electronic, optical, and mechanical properties[1] that make them promising for the realization of field effect transistors[2], integrated photonics devices[3], solar cells[4] and photoelectrochemical cells[5]. GaAs NWs are typically grown by Vapor-Liquid-Solid (VLS) method[6] by Molecular Beam Epitaxy (MBE). Thanks to their small footprint, elastic relaxation occurs at the first stages of the NW growth allowing them to prevent dislocations even on highly mismatched substrate. Thus, GaAs NWs were successfully grown on low-cost Si substrate[7], allowing integration of light sources on the dominant material in microelectronic. GaAs NWs are typically grown by Vapor-Liquid-Solid (VLS) method[6] by Molecular Beam Epitaxy (MBE). In this respect, gold, which is the most common catalyst for VLS growth, can endanger the optoelectronic performances by diffusing and inducing non-radiative defects. Consequently in 2008, Colombo *et al.* proposed to use Ga as a catalyst, instead of Au, and therefore promoted the self-catalyzed VLS growth[8,9]. Usually, GaAs NW self-catalyzed growth is performed directly on epi-ready Si substrates without any removal of the native $SiO_x$ (x<2) overlayer[9,10], but the oxide layer has to be thin enough for Ga to reach the Si surface by drilling holes through $SiO_x$ reduction. It should be noted that self-catalyzed growth of GaAs NWs has been also reported on substrate with thicker $SiO_2$ layer but patterned [11,12]. In this case, it has been shown that the hole morphology is an important feature because it prompts the verticality or non-verticality of the NWs. Besides, the oxide thickness, roughness and stoichiometry have been shown to modify the NW density widely[13–15]. Slight variations of the oxide characteristics are known to induce completely different NWs even for the same experimental growth conditions.

In a typical sequence of GaAs nanowires self-catalyzed VLS growth, three steps are performed. In a first step, termed as "pre-deposition" below, Ga is deposited on the substrate to form nano-droplets (NDs). These NDs must reach the Si substrate from which crystalline and epitaxial NWs can only be grown. The amount of Ga, the deposition temperature, the Ga flux and the surface preparation are critical parameters that strongly affect the NDs morphology[16,17]. In a second step, termed as "post-deposition annealing", the temperature is increased up to the NW growth temperature. The last step is the NW growth itself, which is obtained by opening the Ga and As shutters of the Molecular Beam Epitaxy (MBE) reactor. It should be noted that if the separation into distinct steps is not necessary to obtain actual NW epitaxial growth, it offers a supplementary degree of control of the NW density and size as recently demonstrated[18]. In short, the control of GaAs NW growth on silicon is almost completely dependent on these early stages, especially the deposited Ga amount, the pre-deposition substrate temperature, and the post-deposition annealing temperature. Recently these early stages of the NW growth have been investigated by several groups using various characterization tools such as scanning electron microscopy and atomic force microscopy (AFM)[18–20]. Despite their interest for measuring the ND contact angle, these techniques do not probe the chemical interaction occurring at the ND/oxide interface and are rarely available as *in situ* characterization directly connected to a MBE reactor. In this context, *in-situ* X-ray Photoelectron Spectroscopy (XPS) is a perfect tool thanks to its surface sensitivity and its ability to detect chemical modifications via core-level shifts.

In this study, we investigate the effect of the growth conditions at each step (during Ga pre-deposition and ND formation during post-deposition annealing and the first stages of NW growth) on the chemistry and morphology of the Ga NDs that induce the subsequent growth of the NWs. A particular attention is drawn on the formation of the Ga oxide which is thought to play a crucial role. In particular, the aim of the present paper is to explore the Ga sticking and dewetting on $SiO_x$ surface as a function of substrate temperature as well as the silica-layer etching by Ga, which enables the Ga NDs to reach the crystalline Si substrate from a chemical point of view. For this purpose, X-ray photoemission (XPS) and Atomic Force Microcopy (AFM) are combined.

2. Experimental

The (111)-oriented epi-ready n-doped silicon substrates were cleaned by 30 s immersion in acetone then 30 s in ethanol solution in order to remove residual and carbon contamination. The substrates were then introduced into the MBE chamber, in a base pressure less than $10^{-9}$ Torr, equipped with a gallium Knudsen cell and a valve cracked As source. The substrate temperature was measured by a thermocouple, which presents a discrepancy of about 50-70 °C below the actual temperature. Equivalent thickness to 3 monolayers (ML) of Ga (~1 nm) was then deposited with a flux of either 0.3 ML/s or 0.5 ML/s. The sample holder was rotated to homogenize the Ga deposition and substrate temperature. Finally, the sample was cooled down to room temperature and directly transferred to the XPS chamber using a conveyor belt under UHV.

XPS measurements were carried out in a VSW spectrometer equipped with a focused monochromatic X-ray source Al Kα (1486.6 eV), in a base pressure less than $10^{-9}$ Torr. The acceptance angle of the hemispherical analyzer is around 3°. The energy resolution of the electron analyzer is 0.5 eV. The different spectra were recorded at normal emission. After checking the sample cleanliness by a survey, Si2p, O1s, Ga2p and Ga3d were recorded. The C1s signal is found negligible for all the samples. The presence of As (As2p and AsLMM) was also systematically searched since the MBE reactor is dedicated to the growth of III-V semiconductors and As can be an unwanted contaminant. Since the deposited Ga amount is very small, we chose to analyze Ga2p rather than Ga3d as its photoionization cross-section is ten times larger than that one of Ga3d.

The substrates are covered by $SiO_x$ and therefore, the spectra are slightly shifted by charge effect. Consequently, all along this paper, we have assumed the $Si2p_{3/2}$ peak from the silicon substrate to be at 99.5 eV and we have shifted the other core levels accordingly[21]. The core levels are fitted by either Voigt/Gaussian or Doniach-Sunjic function depending on their metallic character. The inelastic electron background is taken into account accordingly via a Shirley background. Contrary to Si2p which has a spin-orbit splitting of 0.61 eV, Ga2p presents a large spin-orbit splitting of 26.9 eV, therefore, only $Ga2p_{3/2}$ is shown in this paper. To gain complementary information about the surface morphology, AFM analyses were performed ex-situ using a Veeco CP2 microscope in tapping mode with scan rate between 0.5 and 1.0 Hz.

### 3. Results and Discussion

#### 3.1. Formation of the Ga NDs

In order to study the formation of the NDs as a function of the substrate temperature during the deposition, 3 MLs of Ga were deposited at a substrate temperature running from 260 °C to 610 °C. Fig. 1 shows the AFM images of this series. From these images the ND diameter D, height h and density d were determined. The results are displayed in Fig.2. For the lowest temperature (260 °C), NDs with high density and small diameters are observed. As the substrate temperature increases, this density decreases regularly. Simultaneously, the NDs become bigger when the temperature increases from 260 °C to 460 °C, before becoming smaller beyond 560 °C. At 610 °C, the NDs are not seen anymore.

At low temperature Ga nucleates to form a dense array of NDs. Then as substrate temperature increases, diffusion is enhanced and induces formation of bigger droplets but with lower density until desorption becomes too important and the amount of Ga at the surface strongly reduces [22,23].

$Ga2p_{3/2}$ core levels recorded for these samples are shown in Fig. 3. $Ga2p_{3/2}$ has two components: one at 1117.6 eV attributed to the Ga-Ga bond (fitted by a function Doniach-Sunjic function to take into account its metallic character) and one at around 1120.3 eV attributed to the Ga-O bond (fitted by a Gaussian function to take into account its non-metallic character). The very small amount of Ga does not allow to determine accurately the oxidation degree of gallium, but our data are in good agreement with $Ga^{3+}$ [24]. As both Ga deposition and XPS analysis have been done in-situ under UHV, we can conclude that the oxygen needed for Ga oxidation is locally pumped from the $SiO_x$ overlayer. Fig.4a and b show the ratio of $Ga2p_{3/2}$ total area to $Si2p$ total area and the ratio of metallic Ga to Ga total area respectively. The intensity of $Ga2p_{3/2}$ core level decreases with the substrate temperature until it reaches quasi-zero at 610 °C. These results are in complete agreement with AFM results shown in Fig.1 where density of NDs decreases with temperature. At all substrate temperatures the $Ga2p_{3/2}$ core levels show an oxidized component. From 260 to 560 °C, the highest component remains the metallic one. It is only when 610 °C is reached that the oxidized component becomes dominant. Moreover, at this temperature, both the Ga oxide and the metallic Ga components are slightly shifted (0.4 eV) towards higher binding energy (Fig.3), which tends to indicate a different chemical state. This point will be discussed in the next section. Finally, note that the amount of gallium which is still detected by XPS is very low although no NDs are seen by AFM anymore. This difference between XPS and AFM could be explained by their different probed area or by the fact that Ga oxide can be spread over $SiO_x$ uniformly (or even diffused into), and thus can be undetectable by AFM.

Actually, as the density and size of NDs vary with pre-deposition substrate temperature, this latter appears as a handy experimental parameter to control the density and the morphology of the NDs. This tuning of the ND morphology through the growth parameters has been recently investigated in details by Tauchnitz *et al.* and has been proven to be an effective way to control the subsequent NW growth[18].

### 3.2. Post deposition annealing

In this part we have replicated the early stages of standard NW growth conditions[13,14,25]: Ga was deposited at a substrate temperature of 510 °C and the substrate temperature was then increased to reach the temperature at which GaAs NWs are grown. This post-deposition annealing has been performed between 560 °C and 660 °C. The annealing time was fixed at 2 min and each experiment corresponded to a new sample (non-cumulative annealing).

AFM measurements (cf. supplementary information) show that Ga NDs desorption depends on the post-deposition temperature. In a recent study, using conductive AFM, Tauchnitz *et al.*[18] have shown the formation of nano-holes induced by the Ga NDs desorption. These holes were not clearly observed in our AFM images (see supplementary information, Fig. S1) because of rapid Si oxidation or insufficient lateral resolution[18]. But some smaller NDs are still seen at the surface.

The corresponding XPS spectra after post-deposition annealing are shown in Fig.5. The ratio of $Ga2p_{3/2}$ total area to Si2p total area, the ratio of metallic Ga to Ga total area and the shift between metallic Ga peak and oxidized Ga peak are reported in Fig.6. Part of the total Ga amount is desorbed during the post-deposition annealing; the higher the temperature, the higher the desorption rate. Besides, the contribution of the oxidized component in $Ga2p_{3/2}$ is higher than the one of metallic Ga for all the post-deposition annealed samples, unlike for the non-annealed reference sample. Below 600 °C (not shown here) both components are at the same binding energy as before post-deposition annealing but above 600 °C the component attributed to metallic gallium is gradually shifted by up to 0.6 eV towards higher binding energy (see Fig. 5).

It should be noted that this chemical shift cannot be attributed to GaAs from As contamination since no As2p core level or As LMM Auger peak were detected by XPS. At this point, it becomes hazardous to assume a truly metallic behavior for the contribution attributed to Ga-Ga interaction. In this respect, the full width at half maximum (FWHM) of this component increases with the temperature denoting a complex chemical interaction between Ga and the silicon substrate. In particular, one can advance the increasing solubility of silicon in liquid gallium with temperature[26].

Wright and Kroemer[27] have shown that an atomic Ga beam can reduce the SiO$_2$ layer around 800 °C by oxidized itself following these reactions:

4 Ga + SiO$_2$ → 2 Ga$_2$O + Si

2 Ga + SiO$_2$ → Ga$_2$O + SiO

Both Ga$_2$O, SiO and the excess of atomic Ga are volatile at this temperature. Based on these reactions and our results, we can assume that liquid Ga NDs can reduce the SiO$_x$ locally at a lower temperature (around 610 °C) than a Ga molecular beam flux (see supplementary information, Fig. S2).

In short, our results are compatible with two mechanisms: during the post-deposition annealing, liquid Ga NDs drill nano-holes into the SiO$_x$ overlayer by reducing SiO$_x$ and therefore oxidizing itself. The shift of the metallic Ga2p peak above 600 °C appears as a XPS signature of the nano-holes formation and for the fact that NDs Ga have reached the Si substrate.

### 3.3. Nanowires Growth

In order to understand the oxidized-Ga role in the NW growth we have grown GaAs NWs with the pre-deposition of 1ML of Ga at 510 °C followed by annealing at 610 °C. Three samples were fabricated: Sample 1 is the same as previously studied, i.e. Ga deposition plus post-deposition annealing. On Sample 2, very short NWs were grown during 1 min 20 s in order to analyze the first stages of the NW growth. On Sample 3, longer NWs were grown during 35 min. In both cases, Ga and As fluxes were set respectively to 1.4 Å/s and 3.2 Å/s (quoted in units of equivalent growth rates of GaAs 2D layers measured by RHEED oscillations on a GaAs substrate[28]), corresponding to a V/III flux ratio equal to 2.3. The SEM images of samples 2 and 3 are shown in Fig.7. From these images the estimated NW length is 76 ± 19 nm and 3.1 ± 0.5 μm for sample 2 and 3, respectively. Note that the Ga droplet is still observed at the NW tip for both samples 2 and 3 because the As shutter was closed during the cooling right after the end of the NW growth and before the Ga drops could consume. Moreover, a GaAs layer made of coalesced GaAs islands is visible in between the NWs for sample 3, but not for Sample 2 due to the short growth time.

The Ga2p$_{3/2}$ core levels are reported in Fig.7 Si2p can be still used for binding energy reference because the GaAs NWs and 2D are not dense enough to screen the substrate completely. Sample 1 displays the same results as previously described, with the shift of the Ga-Ga component attributed to the ND/substrate interaction. For sample 2 with short NWs, three components are detected and attributed to Ga-O, Ga-As and Ga-Ga. The Ga-Si peak is not seen anymore because the NW growth has concealed-the interaction between Ga and the substrate. The Ga-Ga component arises from the Ga droplets present at the top of the NWs while the Ga-As component come from NWs and GaAs layer. The Ga-O component remains from the Ga oxidation during post-annealing. For sample 3 with long NWs, only one intense peak is visible and therefore attributed to Ga-As both from the NWs and the GaAs layer. It is normal that the large amount of GaAs screens the contribution from Ga-O. However, Ga droplets are still visible at the NW tips on SEM images, but their fingerprint cannot be found in XPS spectra because their contribution is too small to be detected. From these results, we can conclude that even if gallium oxides are still present at the beginning of the NW growth, they do not inhibit the GaAs NWs growth.

4. **Conclusion**

To summarize, our results show that splitting the whole process into three steps, i.e. Ga deposition, post-deposition annealing and NW growth, is a handy approach to control the size, density and growth of the Ga NWs on $SiO_x$/Si(111). First, during Ga pre-deposition, the substrate temperature is a key parameter. The size and the density of the NDs are the result of a balance between Ga diffusion and Ga desorption. Part of the deposited Ga is oxidized because of an oxygen migration from the $SiO_x$ overlayer. At this point the Ga oxides contribute to the droplet self-organization by tuning Ga diffusion on the surface. Second, during the post-deposition annealing needed to reach the NW growth temperature, the Ga oxidation plays a key role as it locally consumes the $SiO_x$ overlayer and drills nano-holes into $SiO_x$ until reaching the Si substrate. This ultimate state is clearly marked in XPS spectra as a shift towards higher binding energy of the Ga2p core level. At this point, the $SiO_x$ overlayer acts as a mask patterned by the Ga droplets which have been self-organized during the previous step. Consequently, the substrate temperature is clearly determinant because it affects the efficiency of redox reaction responsible of the holes. Third, when NWs are grown on this patterned substrate, GaAs grows either as a 2D layer (in between holes on Ga oxide) or as NWs (in holes where there Ga liquid NDs, which can catalyze NW growth , are directly in contact with the crystalline Si surface). Therefore, both Ga oxide and Ga doped Si are buried under GaAs. As a result, in Ga2p core level, the Ga-O component disappears and the metal Ga2p component regains its classical location as its chemical shift before growth is attributed to Ga-Si interaction. On the basis of these results, it is now clear why any chemical or morphological modification of the silica overlayer under a given growth condition affects the NW growth, since it strongly modifies Ga surface diffusion and redox reaction.

**Acknowledgements:**

This work has been partly funded by HETONAN (ANR-15-CE05-009). The authors thank NanoLyon for access to equipment and J. B. Goure for technical assistance. M. G. Silly and M. Gendry are acknowledged for fruitful discussions.

FIGURES:

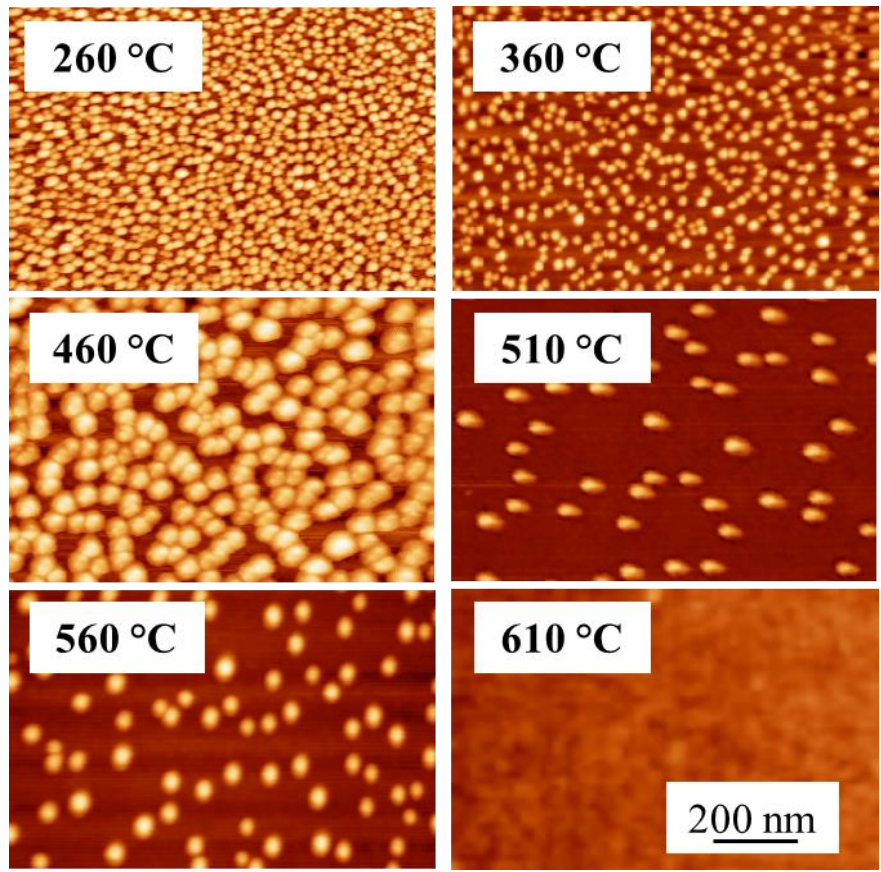

Fig.1 AFM images of 3 monolayers of Ga deposited at different substrate temperatures (from 260 °C to 610 °C).

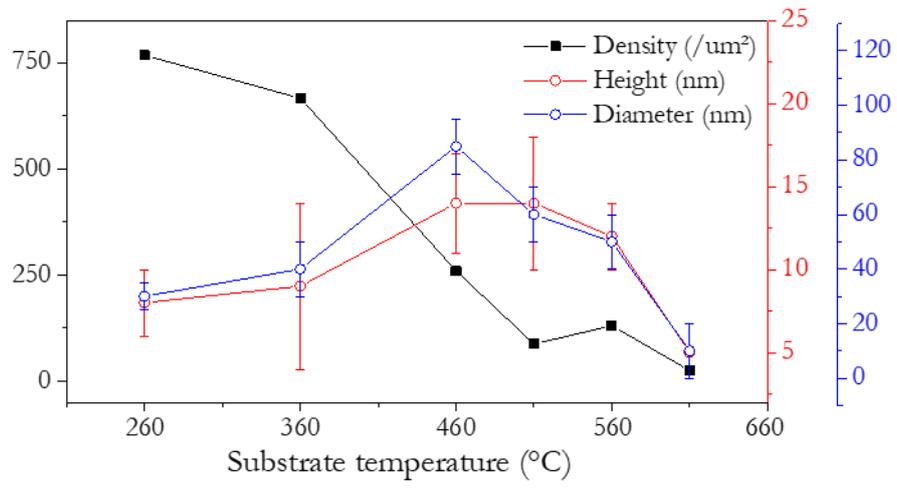

Fig. 2  Ga droplets density (number of droplets/μm²), diameter and height as a function of the substrate temperatures (from 260 °C to 610 °C).

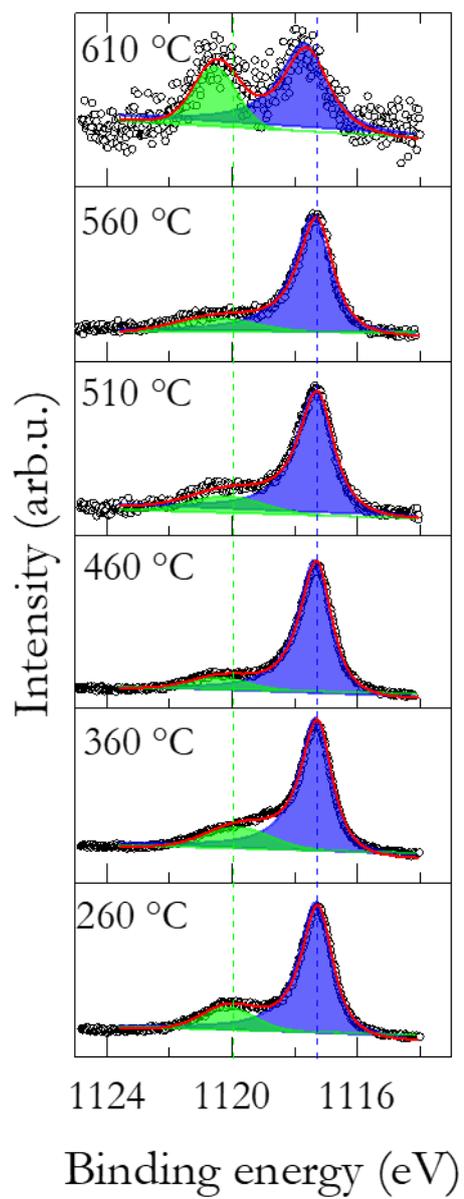

Fig. 3 Ga2p3/2 core level of 3 monolayers of Ga deposited at different substrate temperatures (from 260 °C to 610 °C), fitted with two components: Ga-Ga (metallic Ga) in blue and Ga-O in green; raw data in black; fitted curve in red, binding energy with respect of Si2p at 99.5 eV.

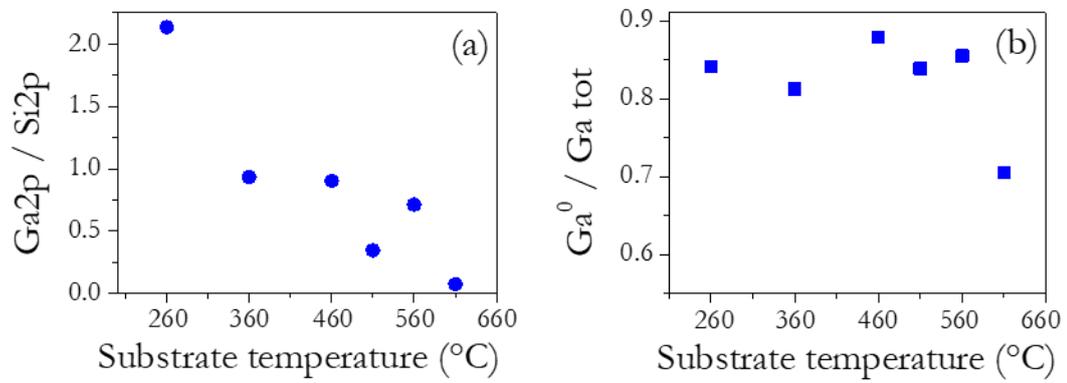

Fig. 4  (a) Ratio of Ga2p3/2 total area to Si2p total area; (b) ratio of metallic Ga to Ga total area, as a function of substrate temperature.

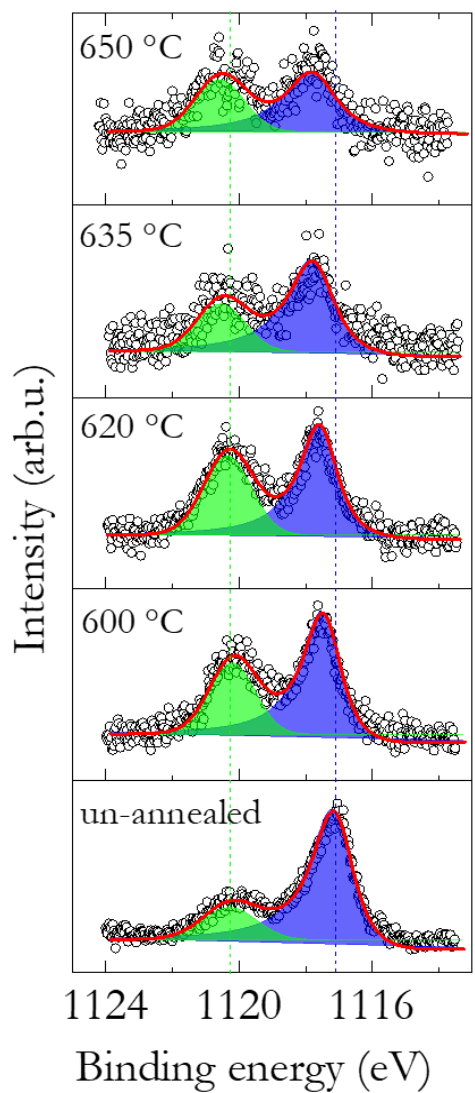

Fig. 5 Ga 2p3/2 core level recorded for Ga NDs deposited at 510 °C and then annealed at different temperatures from 600 to 650 °C; binding energy with respect of Si2p at 99.5 eV.

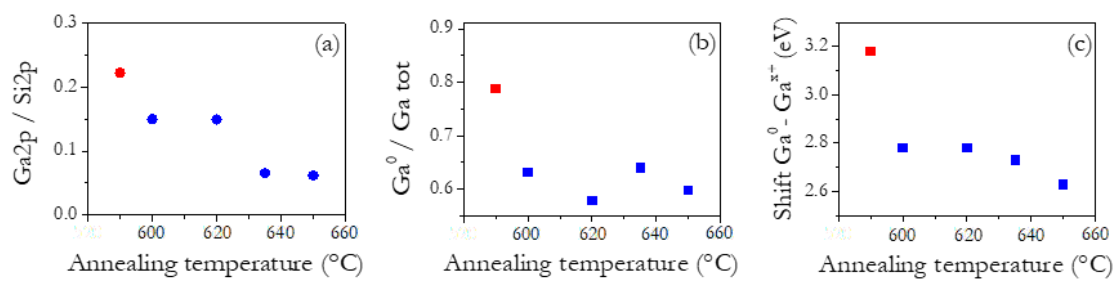

Fig. 6 (a) Ratio of Ga2p3/2 total area to Si2p total area; (b) ratio of metallic Ga to Ga total area; (c) shift between metallic Ga peak and oxidized Ga peak, as a function of annealing temperature. In red the point corresponding to the un-annealed sample.

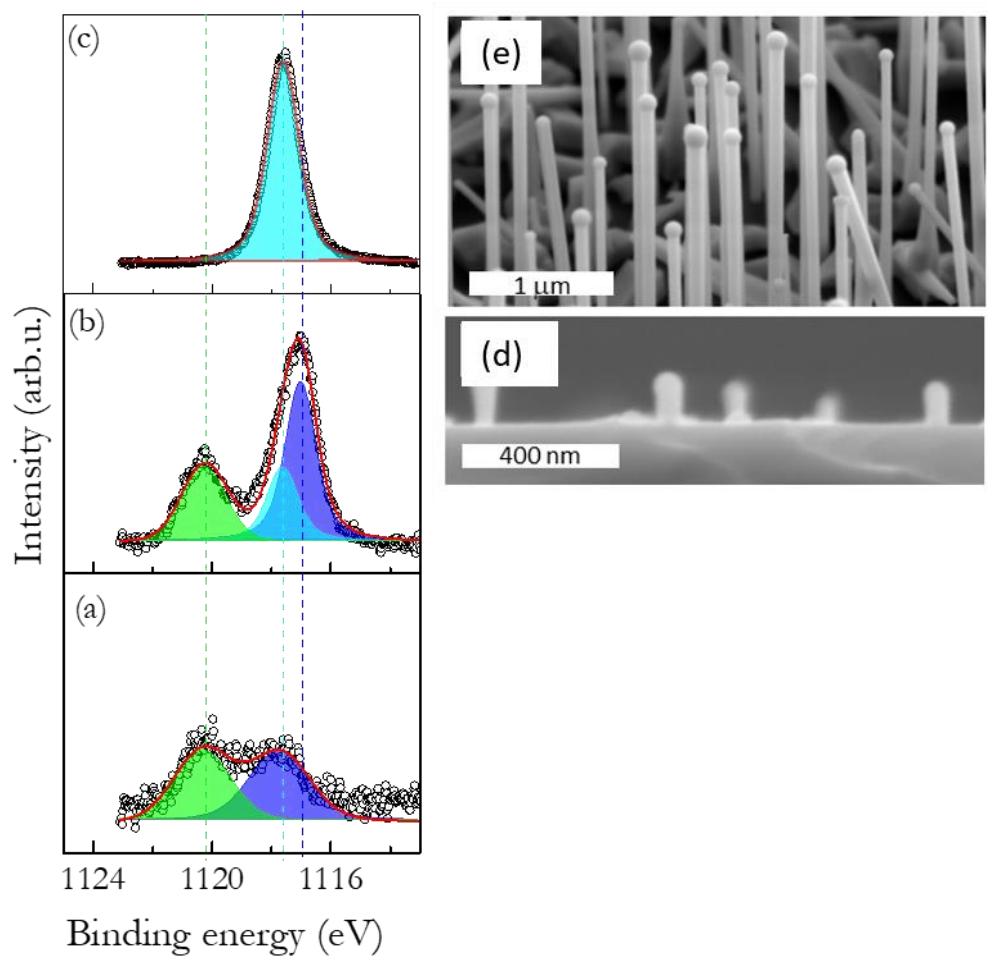

Fig. 7 Ga 2p3/2 core level recorded and SEM images for (a) Sample 1 : Ga NDs deposited at 510 °C and annealed at 610 °C, (b, d) Sample 2: GaAs nanowires 1 min 20 s growth, (c, e) Sample 3 : GaAs nanowires 35 min growth; binding energy with respect of Si2p at 99.5 eV.

Table 1: Core level parameters

| Core Level | Spin Orbit | Cross Section[29] (for AlKα) |
|---|---|---|
| Ga 2p | 26.9 | 0.4412 |
| Ga 3d | 0.46 | 0.014 |
| O 1s |  | 0.04004 |
| Si 2p | 0.61 | 0.26237 |
| As 3d | 0.7 | 0.02517 |
| As 2p | 35.8 | 0.5581 |

Table 2: Measured binding energy

| Core Level | Bonding | Binding Energy (BE) eV |
|---|---|---|
| Ga $2p_{3/2}$ | Ga-Ga | 1117.3 |
| | Ga-O | 1119.7 (aired oxidation) 1119.3 (oxidation on $SiO_x$) |
| | Ga-Si | 1117.7 - 1117.9 |
| | Ga-As | 1117.9 |
| Ga $3d_{5/2}$ | Ga-Ga | 18.75 |
| | Ga-O | 21.15 eV (aired oxidation) ~20 (oxidation on $SiO_x$) |
| | Ga-As | 19.4 |
| Si $2p_{3/2}$ | Si-Si | 99.5 |
| | Si-O | ~103.3 |